\documentclass[twocolumn,superscriptaddress,aps,prl]{revtex4-1}
\usepackage{natbib}
\usepackage{graphicx}
\usepackage{enumitem}
\DeclareGraphicsExtensions{.pdf,.png,.jpg}
\usepackage{array} 
\usepackage{amsmath} 
\usepackage{amssymb}    
\usepackage{latexsym}
\usepackage{amsfonts}
\usepackage{mathrsfs}
\usepackage{color}
\usepackage{hyperref} 
\usepackage{subfigure}

\begin{document}

\newcommand{\bra}[1]{\left< #1\right|}   
\newcommand{\ket}[1]{\left|#1\right>}
\newcommand{\abs}[1]{\left|#1\right|}
\newcommand{\ave}[1]{\left<#1\right>}
\newcommand{\Tr}{\mbox{Tr}}
\renewcommand{\d}[1]{\ensuremath{\operatorname{d}\!{#1}}}

\title{Irreducibility of coherent states}


\author{Jeongwoo Jae}
\affiliation{Department of Physics, Hanyang University, Seoul 133-791, Korea}

\author{Kang Hee Seol}
\affiliation{Department of Physics, Hanyang University, Seoul 133-791, Korea}
\affiliation{Research Institute for Natural Sciences, Hanyang University, Seoul 133-791, Korea}

\author{Kwang-Geol Lee}
\email{kglee@hanyang.ac.kr}
\affiliation{Department of Physics, Hanyang University, Seoul 133-791, Korea}
\affiliation{Research Institute for Natural Sciences, Hanyang University, Seoul 133-791, Korea}

\author{Jinhyoung Lee}
\email{hyoung@hanyang.ac.kr}
\affiliation{Department of Physics, Hanyang University, Seoul 133-791, Korea}
\affiliation{Research Institute for Natural Sciences, Hanyang University, Seoul 133-791, Korea}
\begin{abstract}
{We suggest an optical method which tests a nonclassical feature with a coherent state input. The test is designed with a multiplexer of on/off detectors and post-selection, adopting sub-binomiality as a nonclassical feature, replacing Mandel's $Q$-factor. The sub-binomiality is shown negative even for coherent states when the post-selection is made. However, we show that it can be reproduced also by a classical model assuming a stochastic on/off detectors. In the sense, the sub-binomiality is unlikely to identify the genuine nonclassicality. On the other hand, we propose a coincident probability of first two branches of the multiplexer and show that the classical model fails to reproduce the quantum coincident probability. The failure of the classical model results from the classical description of light, i.e. the divisibility of intensity into parts no matter how small it is. Then our optical test identifies a nonclassical feature of coherent states against the classical divisibility of light, which we call irreducibility.}
\end{abstract}

\maketitle
{\em Introduction.---}Laser is the most fundamental and versatile resource in modern optics~\cite{Zang1990}. Generating coherent states by laser has become the most basic and significant procedure in experimental studies. Coherent states are also responsible for essential parts in quantum-optical experiments and optical realizations of quantum information processings~\cite{Bachor2004}. For instance, they have been adopted to produce nonclassical states such as squeezed states~\cite{Loudon1987}, photon-number states~\cite{Eisaman11,Bocquillon09}, and photon-added states~\cite{Zavatta04, Zavatta09}. Those nonclassical states are requisite elements to carry out quantum information processing; quantum cryptography~\cite{BB84,Stucki05}, quantum teleportation~\cite{Furusawa98,Lee00a} and quantum computation~\cite{Knill01,Jeong2002}.
                                                                                                                                        
The coherent states themselves are commonly regarded to be the most classical of quantum states~\cite{Glauber63}. They have counterparts in the classical theory of electromagnetic fields if replacing the quantum uncertainties with stochastic ones~\cite{Titulaer65,Mandel65,Vogel2000}. Existence of classical counterparts has been identified by Glauber-Sudarshan $P$ function~\cite{Glauber63,Titulaer65,Mandel86} on optical phase space,
\begin{equation}
\hat{\rho} = \int d^2\alpha P(\alpha) |\alpha\rangle\langle\alpha|, \nonumber
\end{equation}
where \{$\ket{\alpha}$\} is an overcomplete basis of coherent states. And states whose $P$ function is a probability measure is said as classical~\cite{Hillery1985, Mandel86}. But $P$ functions of pure states are in general highly singular and difficult to treat in experiments, and the tests for identification have been proposed in terms of marginal probabilities~\cite{Vogel2000} and moments~\cite{Shchukin2005}. Mandel's $Q$-factor also identifies a nonclassical feature of sub-Poisson statistics in the photon-number distribution~\cite{Mandel65,Agarwal92} (which may be regarded as a special case of using moments~\cite{Shchukin2005}). These have been refined recently~\cite{Ryl15,Sperling2012a,*Sperling2012}. They claim that coherent states have the classical counterparts, i.e. coherent states are classical roughly speaking, witnessed by their regular $P$ functions, i.e., probability measures. This is the case no matter how small the intensities are~\cite{Interestingly}.

\begin{figure}[t]
	\centering
	\includegraphics[width=0.38\textwidth]{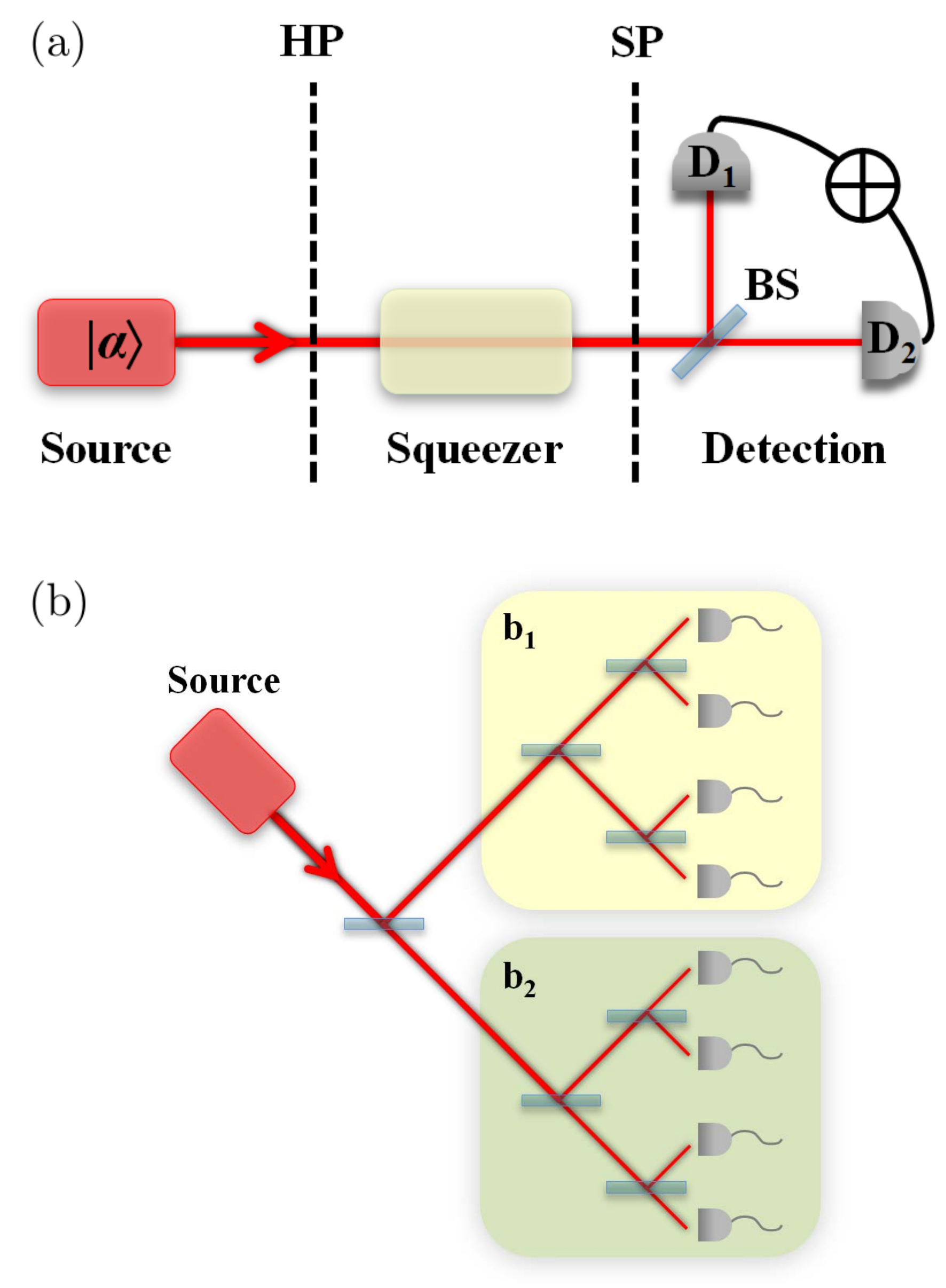}
	\caption{(a) Schematic diagram of {$g^{(2)}$} test with a squeezed coherent state. The inclusion of squeezing operation determines the presence of the nonclassical feature. It belongs to the state preparation or the measurement, depending on the picture; Schr{\"o}dinger picture (SP) or Heisenberg picture (HP), respectively. $D_{1,2}$ are ideal photon counters, placed after a beamsplitter (BS). (b) Realization of $g^{(2)}$ test by using two PNRCs (photon number-resolving counters) to test a nonclassical feature with a coherent state input. BSs are placed at joints of modes. We make the post-selection that detection events are selected if they are of single or no click at each branch $b_{1,2}$.}
	\label{fig:setting}
\end{figure}

However, a nonclassical feature is not always the consequences that a state characterizes, because each test result is produced by both of state and measurement. By `a nonclassical feature', thus, we mean rather \textit{a quantum prediction that a classical model is unable to reproduce}. For light fields, two types of nonclassical features have mainly been discussed in literatures; each against the classical theory of electromagnetic fields~\cite{Glauber63,Vogel2000,Lee00a} or against hypothetical theories of hidden variables~\cite{Brunner2014,Banaszek1998,Lee00c}. Every quantum prediction results from the interplay of a quantum state and a measurement, so that it can inherit a nonclassical feature from the one(s) of a state~\cite{Glauber63,Titulaer65,Mandel65,Mandel86,Lee00a} and/or a measurement~\cite{Banaszek1998,Lee00c,Chen2002}. For instance, consider a test of the 2nd order coherence $g^{(2)}$ where a squeezer intervenes the preparation of a coherent state. The test shows a nonclassical feature of anti-bunching. Adopting Schr{\"o}dinger picture, one may argue that the squeezed coherent state has non-regular $P$ function and the state originates the nonclassical feature~\cite{ScullyCH4}. On the other hand, adopting Heisenberg picture, it may be argued that the initial coherent state has a regular $P$ function and the nonclassical feature originates from the measurement accompanying the squeezer [see Fig.~\ref{fig:setting}(a)]. This example shows that the measurement can be solely responsible for the nonclassical feature, while the source is classical~\cite{One}. In addtion, it is interesting that coherent states can have negative values in unconventional number-phase Wigner functions~\cite{Vaccaro1995} and they occasionally result in negative weak values of photon number in a weak measurement~\cite{Johansen2004}. Thus, it is still misty whether coherent states have own nonclassical features that are distinct from those of measurements.

In this letter, we suggest an optical method which tests a nonclassical feature with a coherent state input. The test is designed with a multiplexer of on/off detectors and post-selection. Detection events are selected if they are of single or no click at each branch. We adopt the sub-binomiality as a nonclassical feature, replacing Mandel's $Q$-factor, to eliminate any fictitious effect with multiplexing of on/off detections~\cite{Sperling2012,Sperling2012a}. The sub-binomiality is negative even for coherent states when the post-selection is made. This nonclassical feature might be argued to inherit from the measurement accompanying the post-selection. However, we show that it can be reproduced also by a classical model assuming a stochastic on/off detectors, that are assumed to fire only by the larger intensity than a threshold. In the sense, the sub-binomiality is unlikely to identify the genuine nonclassicality. We consider, on the other hand, a coincident probability (CP) of two branches of multiplexer ($b_1$ and $b_2$ in Fig.~\ref{fig:setting}(b)) and show that the classical model fails to reproduce the quantum CP. In particular, as increasing the multiplexing degree, the classical CP always decreases to zero, suddenly at a certain degree of multiplexing, whereas the quantum CP is saturated to a finite value. The failure of the classical model results from the classical description of light, i.e. the divisibility of intensity into parts no matter how small it is. Thus our optical test identifies a nonclassical feature of coherent states against the classical divisibility of light, which we call `irreducibility'.

{\em Optical test with post-selection.---} We employ a multiplexer of avalanched photodiode (APD) detectors of Geiger mode, which was proposed as a feasible photon number-resolving counters (PNRC). The APD detectors are assumed to distinguish only the presence of photon(s) by a click signal~\cite{Eisaman11}, called on/off detectors. To obtain full anatomy of an incident beam, PNRCs have been proposed by using time delayed multiplexing~\cite{Fitch03} and spatial arrays of on/off detectors~\cite{Jiang07}. Our optical test of a multiplexer consists of beamsplitters (BS), suggested in Ref.~\cite{Sperling2012}. BSs of 50:50 are placed to distribute the incident beam as depicted in Fig.~\ref{fig:setting}(b). The number of final output modes $N$ is called the degree (size) of multiplexing with $N=2^m$ for a multiplexer of depth $m$. For a given input beam, statistics are characterized by the joint distribution of click events at $N$ on/off detectors. In particular, we are interested in the correlation between the two branches $b_1$ and $b_2$ in Fig.~\ref{fig:setting}(b) and obtain the two-branch joint probability $\text{Pr}^0(k_1,k_2)$ instead of the full distribution, where $k_i$ is the number of total clicks by on/off detectors at branch $b_i$. When we make the post-selection of neglecting multi-click events (i.e. $k_i\geq2$), the two-branch joint probability $\text{Pr}^0(k_1,k_2)$ is modified to $\text{Pr}(k_1,k_2)$ with the normalization condition ${\text{Pr}}(1,1)+{\text{Pr}}(1,0)+{\text{Pr}}(0,1)+{\text{Pr}}(0,0)=1$.

We employ two measures to identify nonclassical features. One is sub-binomiality $Q_B$, defined by~\cite{Sperling2012}
\begin{equation}
Q_B = N\frac{\langle(\Delta k)^2\rangle}{\langle k\rangle(N-\langle k \rangle)}-1,
\end{equation}
where $\langle k\rangle$ is the average number of the total clicks and $\langle(\Delta k)^2\rangle$ is the variance. For two branches as in our case, $\langle k\rangle=\sum_{k_1,k_2}(k_1+k_2)\text{Pr}(k_1,k_2)$, and $\langle(\Delta k)^2\rangle=\sum_{k_1,k_2}(k_1+k_2-\langle k\rangle)^2\text{Pr}(k_1,k_2)$. The sub-binomiality was recently proposed to identify intensity fluctuation of source with a PNRC, replacing Mandel's $Q$-factor, $Q_M$ as incorrectly witnessing the presence of nonclassicality due to a ficticious effect of  multiplexer~\cite{Sperling2012}. It becomes negative when the source shows the sub-binomial characteristics in the click distribution or a non-regular $P$ fucntion. This sub-binomiality is a signature for nonclassicality. It was also shown that the sub-binomiality $Q_B$ converges to $Q_M$ in the limit of large multiplexing degree, $N\rightarrow\infty$. The other measure is the coincident probability,
\begin{equation}\label{eq:CP}
\text{CP}=\sum_{k_1,k_2\neq0} \text{Pr}(k_1,k_2).
\end{equation}
This measures the correlation of simultaneous clicks in two branches, $b_1$ and $b_2$~\cite{Eisaman11,Paul1996}. This becomes unity, CP$=1$, in the strong-field limit of incident beam. For a weak field, CP is less than 1.

{\em Quantum model.---} Even though it is a quantum state, a coherent state shares some properties with classical light. In particular the input amplitude is divided into two by a BS operation, $\ket{\alpha}\rightarrow\ket{\alpha_1}\ket{\alpha_2}$ with $\abs{\alpha}^2=\abs{\alpha_1}^2+\abs{\alpha_2}^2$. In the process, no entanglement is generated, as predicted in Ref.~\cite{Springer2009}: BS operation generates no entanglement if the input state has a regular $P$ function. Furthermore, sub-binomiality of a coherent state is non-negative, $Q_B=0$, when no post-selection is made in a multiplexer~\cite{Sperling2012a}. Nevertheless, when making the post-selection that neglects the multi-click events, we show that the sub-binomiality $Q_B$ is negative even for a coherent state input.

The test involves the correlation measurement between the two branches in a mulitplexer, shown in Fig.~\ref{fig:setting}(b). With coherent state input, the click distribution of $k$ clicks is given by the quantum average of the observable~\cite{Sperling2012},
\begin{equation}
\hat{\pi}_{k}=~:{N \choose k}(1-e^{-\frac{\hat{n}}{N}})^{k}(e^{-\hat{n}/N})^{N-k}:, \nonumber
\end{equation}
where ${N \choose k}$ are binomial coefficients, $\hat{n}$ is the number operator, and $:\cdot:$ is normal ordering prescription. Note that a branch may be seen as a multiplexer of degree $N/2$. The composite events of clicks at the two branches are determined by a composite observable, $\hat{\pi}_{k_1}\otimes\hat{\pi}_{k_2}$. The input coherent state $\ket{\alpha}$ is transformed by the first BS into the factorized state, $\ket{\psi}=\ket{{\alpha_1}/{\sqrt{2}}}_1\otimes\ket{{\alpha_2}/{\sqrt{2}}}_2$, where $\abs{\alpha_{1,2}}=\abs{\alpha}$.  The two-branch joint probability $\langle\psi|\hat{\pi}_{k_1}\otimes\hat{\pi}_{k_2}|\psi\rangle$ is 
\begin{figure}[t]
	\centering
	\includegraphics[width=0.44\textwidth]{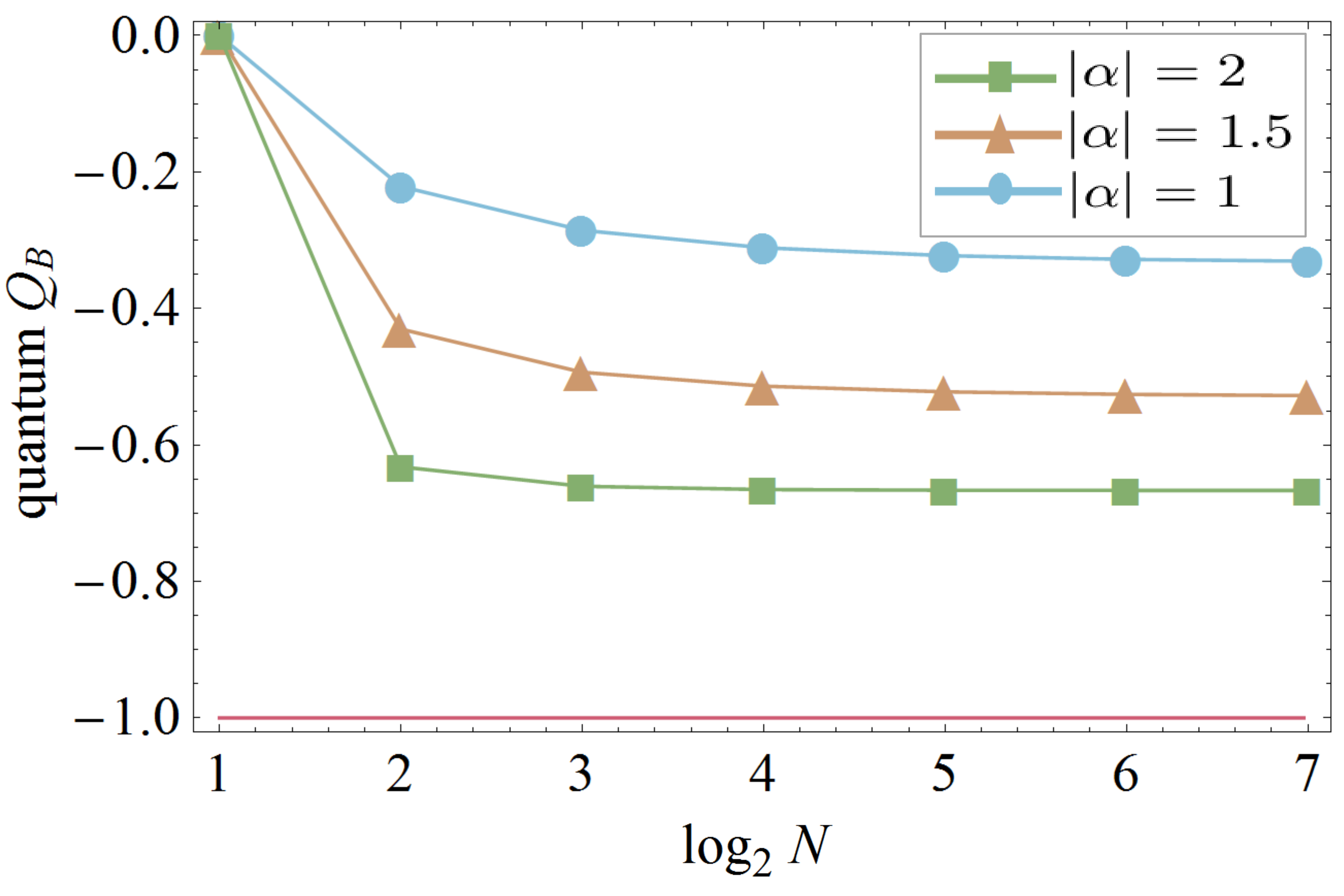}
	\caption{Quantum sub-binomiality $Q_B$ of coherent states as a function of the multiplexing degree in logarithmic scale $\text{log}_2 N$. Coherent state with non-zero amplitudes ($\alpha\neq0$) have negative $Q_B$ when the post-selection is made. The red line of $Q_B=-1$ is of the coherent state with the strong amplitude $\abs{\alpha}=50$. Lines are for eye guides.}
	\label{fig:quantum}
\end{figure}
\begin{equation}\label{eq:bare}
\text{Pr}^{0}(k_1,k_2)=\prod\limits_{i=1,2} {\frac{N}{2} \choose k_i} (c/d)^{k_i}d^{n_i},
\end{equation}
where $c=1-e^{-\abs{\alpha}^2/N}$ is the probability of a detector to click, $d=e^{-\abs{\alpha}^2/N}$ is that of no click, and $c/d$ is the odds for a detector to click. The post-selection of choosing no- and single-click events ($k_i=0,1$) modifies the two-branch joint probability $\text{Pr}^{0}$ to Pr with the normalization condition $\sum_{k_1,k_2=0,1}\text{Pr}(k_1,k_2)=1$,
\begin{equation}\label{eq:normal}
\text{Pr}(k_1,k_2)=\frac{C^{k_1+k_2}}{(C+1)^2}.
\end{equation}
where $C=(N/2)(c/d)$ is a branch odds for the click. Note $C \rightarrow C_{\infty}=\abs{\alpha}^2/2$ as $N\rightarrow\infty$.

The sub-binomiality $Q_B$ is obtained using the two-branch joint probability Pr and given by
\begin{equation}
Q_B=-\frac{\left( 1-\frac{2}{N} \right)C}{\left( 1-\frac{2}{N} \right)C+1}. \nonumber
\label{eq:coincident}
\end{equation}
It is remarkable that the sub-binomiality is always negative for all $\alpha\neq0$ and $N>2$ even for the coherent state input, while it vanishes for $N=2$. Furthermore, as increasing the multiplexing degree $N$, $Q_B$ converges down to a negative value $Q^{\infty}_B = -{C_{\infty}}/(C_{\infty}+1)$ and $Q^{\infty}_B\rightarrow-1$ in the strong-field limit $\abs{\alpha}^2\rightarrow\infty$. These properties are illustrated in Fig.~\ref{fig:quantum}. Reminding of $Q_B=0$ for the bare distribution Pr$^0$ with no post-selection, it is tempted to say that the photon-counting measurements accompanying the post-selection presents a nonclassicality even with a coherent state input, in other words, the nonclassicality of sub-binomiality originates from the combined measurement of photon detections and post-selection, but not from the source~\cite{work}. However, a classical model can reproduce the negativity of $Q_B$, as we show.
	
{\em Classical model.---} We employ a classical stochastic model to mimic the quantum model. The classical model assumes classical electromagnetic fields to be divisible into parts no matter how small their amplitudes are. On the other hand, it assumes on/off detectors to output a click signal for the intensity $I$ larger than a certain threshold $I_{\text{th}}$ (one may program to realize those detectors). In this way, our classical model maintains the source to be an infinitely divisible wave, whereas it mimics the quantum detection of a photon that does not fire until receiving a photon or more.
\begin{figure}[b]
	\centering
	\includegraphics[width=0.5\textwidth]{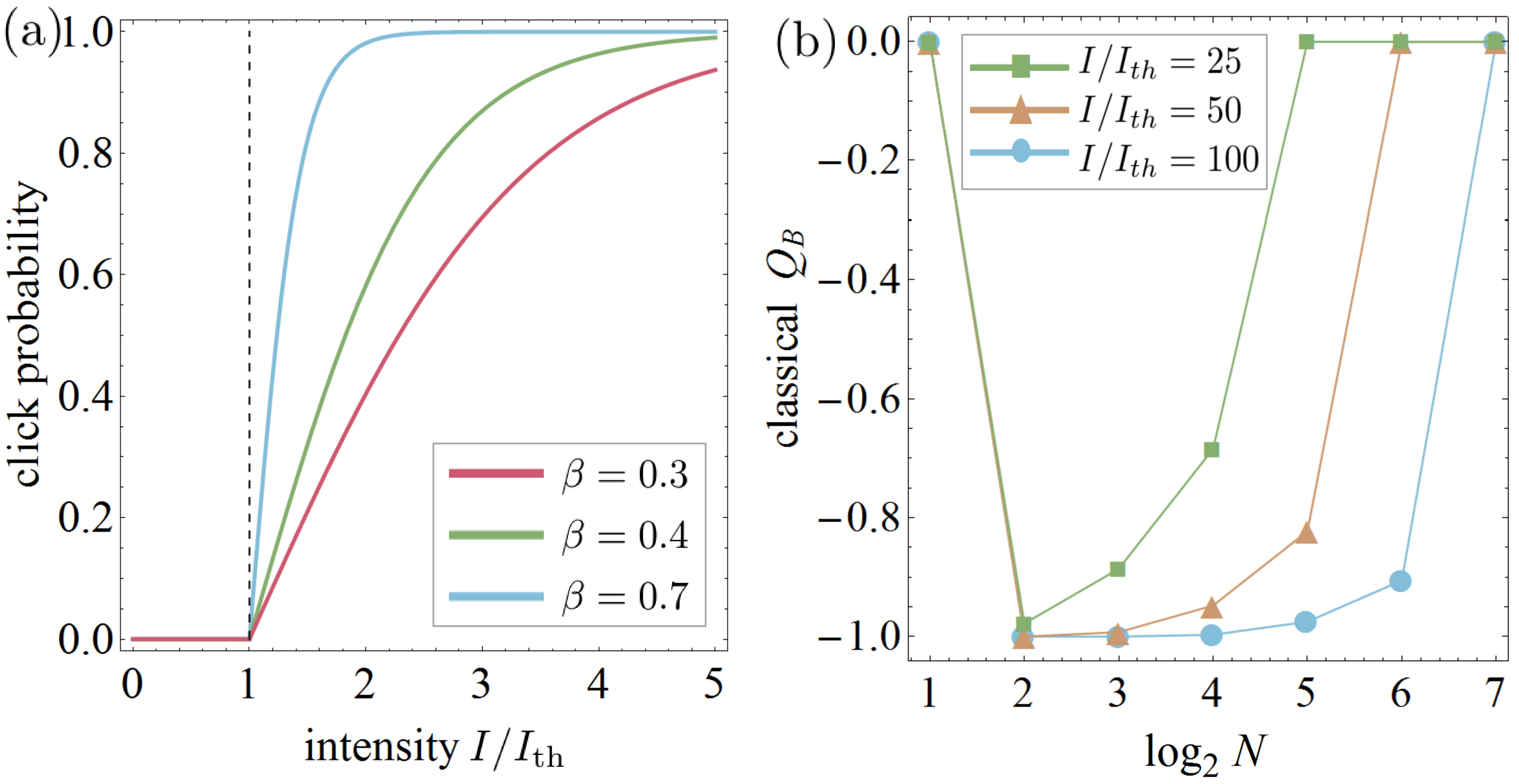}
	\caption{(a) Click probability of a classical on/off detector as a function of input intensity $I$ relative to the threshold $I_{\text{th}}$, for ionization factor $\beta=$ 0.3, 0.4, and 0.7. (b) Classical sub-binomiality $Q_B$ as a function of the multiplexing degree in logarithmic scale $\text{log}_2 N$ for given input intensities. $Q_B$ goes down negative and up to zero as $\text{log}_2 N$ increases. The larger input intensity $I$, the larger sub-binomiality $Q_B$. Lines are for eye guides.}
	\label{fig:classical}
\end{figure}

The classical on/off detector is based on the realistic model of APD detector~\cite{Dautet93}, with its click probability conditioned on an input intensity $I$,
\begin{equation}\label{eq:click}
\text{Pr}(\text{on}|I)=\tanh \left( \frac{ I-I_{\text{th}}}{I_{\text{th}}}f(\beta) \right) \Theta( I-I_{\text{th}}),
\end{equation}
where $\Theta$ is a Heaviside step function. No-click probability is given by, {Pr}$(\text{off}|I)=1-${Pr}$(\text{on}|I)$. The click probability as a function of intensity is presented in Fig.~\ref{fig:classical}(a). The detector fires only when the input intensity is higher than the threshold $I_{\text{th}}$. Here, the stochastic factor $f(\beta)=\beta/(1-\beta)$, where $\beta$ is the degree of ionization of APD, ranging from 0 to 1~\cite{Dautet93}. The value $\beta=0$ implies that the ionization does not occur with $f(0)=0$, while  $\beta=1$ denotes the maximal ionization, $f(1)\rightarrow\infty$. A strong-intensity input always succeeds to click, i.e. {Pr}$(\text{on}|I \gg I_{\text{th}}) \approx 1$. This type of on/off detectors were used in simulating a Bell test~\cite{Diego16}. Our classical multiplexer is assumed to consist of such on/off detectors. In the balanced multiplexer of degree $N$, the input intensity $I$ is uniformly distributed to the output modes with $I/N$.

The classical two-branch joint probabilities, bare $\text{Pr}^0(k_1,k_2)$ and modified $\text{Pr}(k_1,k_2)$, are obtained as in Eqs.~(\ref{eq:bare}) and~(\ref{eq:normal}) by replacing the click and no-click probabilities $c$ and $d$ with the classical $\bar{c}=\text{Pr}(\text{on}|I/N)$ and  $\bar{d}=\text{Pr}(\text{off}|I/N)$, respectively. With no post-selection, the bare distribution shows the binomial statistics, $Q^0_B=0$. This is the same as in the quantum model for coherent states~\cite{Sperling2012a}. With the post-selection that neglects the multi-click events, on the other hand, $Q_B$ becomes negative as if the statistics were sub-binomial or nonclassical, as shown in Fig.~\ref{fig:classical}(b). It is clear that this result, $Q_B<0$, has nothing to do with any nonclassicality, as it comes from all the classical. In the sense, $Q_B$ is unlikely to identify the genuine nonclassicality. Instead, we shall consider another measure of the coincident probability CP in Eq.~(\ref{eq:CP}).

{\em Coincident probability and failure of classical simulation.---} The coincident probability CP is obtained by using the modified joint probabilities Pr$(k_1,k_2)$,
\begin{equation}
\text{CP}=\left(\frac{C}{C+1}\right)^2,
\end{equation}
where the branch odds $C=(N/2)(e^{\abs{\alpha}^2/N}-1)$ for the quantum model and $C=(N/2)\text{Pr}(\text{on}|I/N)/\text{Pr}(\text{off}|I/N)$ for the classical model. We note that CP coincides with the square of $Q_B$, or $Q^{\infty}_B=-\sqrt{\text{CP}_{\infty}}$, in the limit of large multiplexing degree $N\rightarrow\infty$. In the sense, the two measures become equivalent in the limit $N\rightarrow\infty$, regardless of the type of models. 
\begin{figure}[t]
	\centering
	\includegraphics[width=0.45\textwidth]{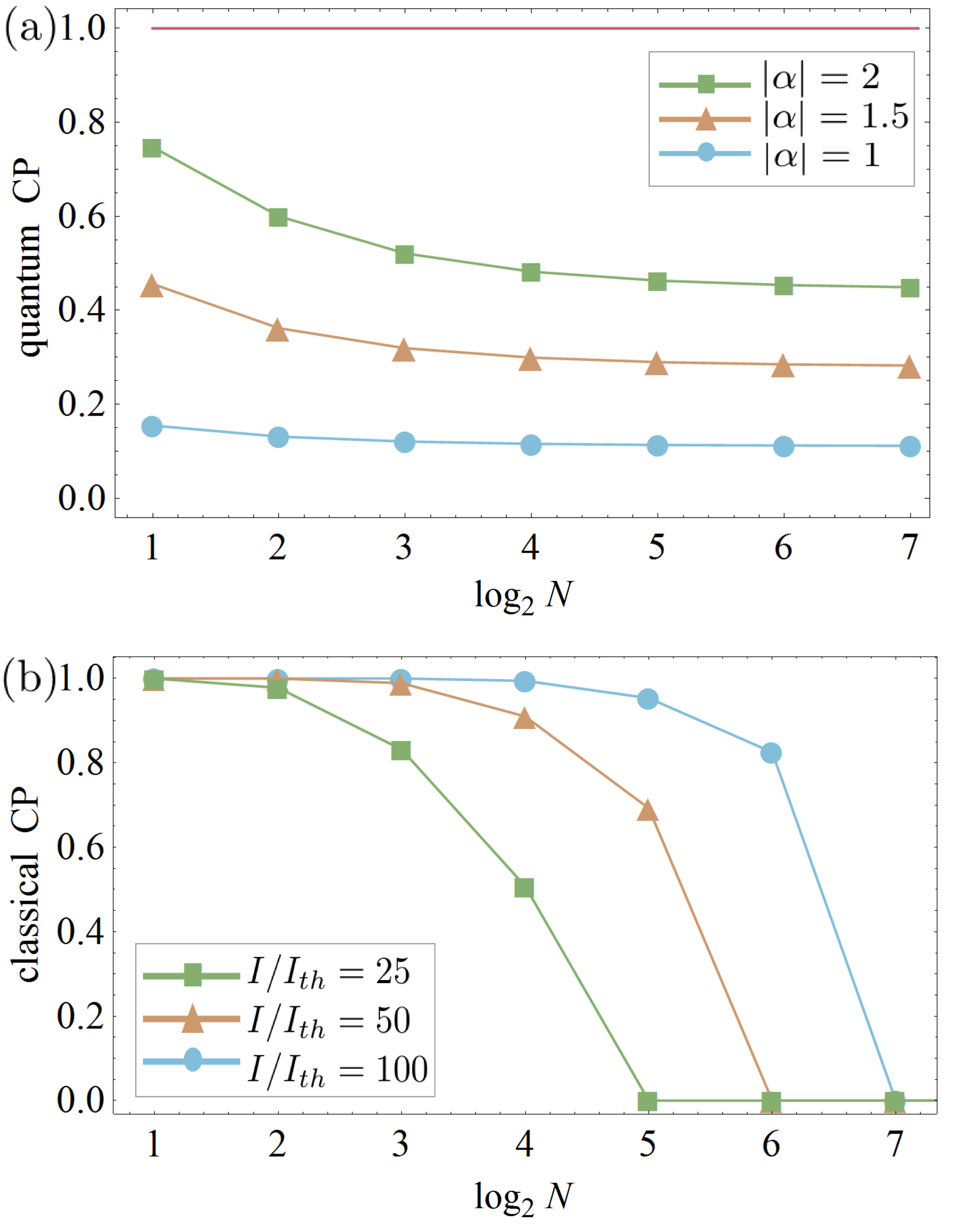}
	\caption{(a) Quantum coincident probability (CP) as a function of the multiplexing degree in logarithmic scale for coherent states. CP$\approx1$ for the strong-amplitude coherent state (red line). For weak-amplitude coherent states, their CP decrease with $\text{log}_2 N$ but saturate to non-zero finite values, depending on the amplitudes $\abs{\alpha} =1$ (circle), 1.5 (triangle), and 2 (square). (b) Classical CP as a function of the multiplexing degree in logarithmic scale. It decreases with increasing the multiplexing degree and suddenly goes to 0 for all input intensities. Given examples are $I=25I_{\text{th}}$ (square), $50I_{\text{th}}$ (triangle), and $100I_{\text{th}}$ (circle) in the unit of $I_{\text{th}}$. Lines are for eye guides.}
	\label{fig:cp}
\end{figure}

The quantum CP decreases as the multiplexing degree $N$ increases. It saturates to a {\em finite} value CP$_{\infty}=(C_{\infty})^2/(C_{\infty}+1)^2$ with $C_{\infty}=\abs{\alpha}^2/2$ for the input coherent state $\ket{\alpha}$. These are illustrated in Fig.~\ref{fig:cp}(a). The classical CP decreases to zero, suddenly at a certain large $N$, as shown in Fig.~\ref{fig:cp}(b), contrary to the quantum. As $N$ increases, the output intensities $I/N$ become less than the threshold, $I/N < I_{\text{th}}$, so that the detectors do not click, as seen in Eq.~(\ref{eq:click}), depending on the ionization degree $\beta$. The classical model thus fails to simulate the asymptotic behavior of the quantum CP. The failure is caused by the classical description of light, in particular, by the assumption of infinite divisibility in our classical model. This result holds even for arbitrarily random classical-distributions of incident beam, as shown in Appendix~A.

The failure of classical simulation implies that the coherent states cannot be described in terms of infinitely divisible fields. The nonvanishing CP results if the division of fields are limited with a certain unit and the unit is not weaker than the threshold intensities of detectors~\cite{Kimble77}. We call the limitation `irreducibility'. One may propose another classical model with the assumption that `beamsplitters' have the limitation of irreducibility, i.e. they are programmed to output the intensities not less than the unit. In particular, the unit may be assumed to be not weaker than the threshold of the classical on/off detectors. Then, the classical model may predict its CP similar to the quantum CP. Thus we do not claim that there exist no classical models which simulate the irreducibility. However, we would claim that the irreducibility is a quantum feature against the classical divisibility on which the classical fields are based.

{\em Conclusion.---} We suggested an optical test which unravels the nonclassical feature of irreducibility with a coherent state input. The test was designed with a multiplexer of the on/off detector and the post-selection. The post-selection chooses the detection events of single or no click at each branch. We show that the sub-binomiality is unlikely to identify the genuine nonclassicality as it is reproduced by the classical model assuming the stochastic on/off detectors. We then proposed another measure of the coincident probability (CP) between the two branches in the multiplexer. It was shown that the classical model fails to simulate the quantum prediction of CP. The failure of the classical model results from the classical description of lights, i.e. the divisibility of intensity into parts. These results imply that our test identifies the nonclassical feature of {\em irreducibility} against the classical divisibility of light and furthermore the coherent states possess the irreducibility.

We thank Prof. M. Plenio for discussion and Prof. W. Vogel for bringing Refs.~\cite{Agarwal92} and~\cite{Ryl15} to our attention. This research was supported by the National Research Foundation of Korea (NRF) grant (No.2014R1A2A1A10050117), funded by the MSIP (Ministry of Science, ICT and Future Planning), Korea government, and it was under the ITRC (Information Technology Research Center) support program (IITP-2016-R0992-16-1017) supervised by the IITP (Institute for Information \& communications Technology Promotion).

\setcounter{equation}{0}
\renewcommand{\d}[1]{\ensuremath{\operatorname{d}\!{#1}}}
\renewcommand{\thesection}{A\arabic{section}}
\renewcommand{\theequation}{A\arabic{equation}}
\section{Appendix A : Vanishing of classical coincident probability for large degree of multiplexing}
In this supplementary material, we show that classical coincident probabilities (CP) of random classical distributions vanish, as increasing the multiplexing degree. Suppose that the incident beam is randomly generated with a probability density function $p(I)$, satisfying
\begin{equation}\label{eq:condition}
\int_{0}^{\infty}dI \, p(I)=1.
\end{equation}
The classical multiplexer is supposed to consist of $N$ on/off detectors and each detector has the click probability, replacing Eq.~(5) with the average over $p(I)$, 
\begin{equation}\label{eq:on}
\text{Pr}(\text{on} | N) =\int_0^\infty dI \, \text{Pr}(\text{on} | I, N) \, p(I) ,
\end{equation}
where $\text{Pr}(\text{on} | I, N) = \tanh \left( \frac{ I/N-I_{\text{th}}}{I_{\text{th}}}f(\beta) \right) \Theta( I/N-I_{\text{th}})$.
The classical CP of the input distribution $p(I)$ is obtained by recasting Eq.~(6),
\begin{equation}\label{eq:term}
\text{CP}=\left( \frac{N\text{Pr}(\text{on}|N)}{N\text{Pr}(\text{on}|N)+2\text{Pr}(\text{off}|N)} \right)^2,
\end{equation}
where $\text{Pr}(\text{off}|N) = 1- \text{Pr}(\text{on}|N)$. It is clear that $\lim\limits_{N \rightarrow \infty}\text{Pr}(\text{on}| N)=0$ and $\lim\limits_{N \rightarrow \infty}\text{Pr}(\text{off}| N)=1$. 
To find the value of $\text{CP}_\infty = \lim \limits_{N \rightarrow \infty}\text{CP}$, we introduce the function $F(x)$,
\begin{equation}
F(x) =\int_{0}^{\infty}dI \, T(x,I) \, p(I) - \int_{0}^{ I_{\text{th}}/x}dI \, T(x,I) \, p(I), \nonumber
\end{equation}
where $T(x,I) = \tanh \left( \frac{ x I-I_{\text{th}}}{I_{\text{th}}}f(\beta) \right)$.
This satisfies $\lim \limits_{x \rightarrow 0} F(x)=0$ and $\lim \limits_{x \rightarrow \infty} F(x)=1$. Then,
\begin{eqnarray}\label{result}
\lim\limits_{N\rightarrow\infty}N \, \text{Pr}(\text{on}|N) &=& \lim_{N \rightarrow\infty} N F(1/N) = \left. \frac{d F(x)}{dx} \right|_{x=0} \nonumber \\
&=& \lim_{N\rightarrow \infty}~  I_{\text{th}} N^2 p( I_{\text{th}} N)T(1/N, I_{\text{th}}N) \nonumber \\
&=& 0,
\end{eqnarray}
where we used $T(1/N, I_{\text{th}} N) = 0$. This remains valid for arbitrary $p(I)$. Thus, the classical CP vanishes, $\text{CP}_\infty = 0$, in the limit of $N \rightarrow \infty$ for arbitrary random distributions of the incident classical beams.

\bibliographystyle{apsrev4-1}

\begin{thebibliography}{99}
\bibitem{Interestingly} Interestingly, Mandel argued that the state attenuated to weak intensity is nonclassical~\cite{Mandel86}.
\bibitem{One} One may take another view that the operation of squeezing is responsible for the nonclassical feature. We would emphasize here the vagueness in claiming what part of a process from the preparation to the measurement originates the nonclassical feature.
\bibitem{work} One may make an argument to find another explanation of a nonclassical source, by moving the effect of the post-selection to that of state engineering. This is beyond this work.
\end{thebibliography}

\end{document}